\documentclass[aps,reprint,superscriptaddress,showpacs]{revtex4-1}
\usepackage{color}

\usepackage[dvips]{graphicx}
\usepackage[dvips]{graphics}

\begin{document}

\title{a-b anisotropy of the intra-unit-cell magnetic order in $\rm YBa_2Cu_3O_{6.6}$}

\author{Lucile Mangin-Thro$^*$}
\affiliation{Laboratoire L{\'e}on Brillouin, CEA-CNRS, Universit\'e Paris-Saclay, CEA Saclay, 91191 Gif-sur-Yvette, France}

\author{Yuan Li$^\#$}
\affiliation{Max Planck Institute for Solid State Research, 70569 Stuttgart, Germany}

\author{Yvan Sidis}
\affiliation{Laboratoire L{\'e}on Brillouin, CEA-CNRS, Universit\'e Paris-Saclay, CEA Saclay, 91191 Gif-sur-Yvette, France}

\author{Philippe Bourges}
\affiliation{Laboratoire L{\'e}on Brillouin, CEA-CNRS, Universit\'e Paris-Saclay, CEA Saclay, 91191 Gif-sur-Yvette, France}

% $^*$ e-mail: mangin-throl@ill.fr\\
% $^\#$ e-mail: philippe.bourges@cea.fr\\

\date{\today}
\begin{abstract}

Within the complex phase diagram of the hole-doped cuprates, seizing the nature of the mysterious pseudo-gap phase is essential to unravel the microscopic origin of high-temperature superconductivity. Below the pseudo-gap temperature $\rm T^{\star}$, evidences for intra-unit-cell orders breaking the 4-fold rotation symmetry have been provided by neutron diffraction and scanning tunneling spectroscopy.  Using polarized neutron diffraction on a detwinned 
$\rm YBa_2Cu_3O_{6.6}$ sample, we here report a distinct a-b anisotropy of the intra-unit-cell magnetic structure factor below $\rm T^{\star}$, highlighting that intra-unit-cell order in this material breaks the mirror symmetry of the CuO$_2$  bilayers. This is likely to originate from a crisscrossed arrangement of loop currents within the $\rm CuO_2$ bilayer, resulting in a bilayer mean toroidal axis along the $\rm {\bf b}$ direction. 

\end{abstract}

%72.15.Eb	Electrical and thermal conduction in crystalline metals and alloys
%74.72.Gh	Hole-doped
%74.72.Kf	Pseudogap regime

\maketitle
%\newpage

\begin{figure}
\includegraphics[width=8cm]{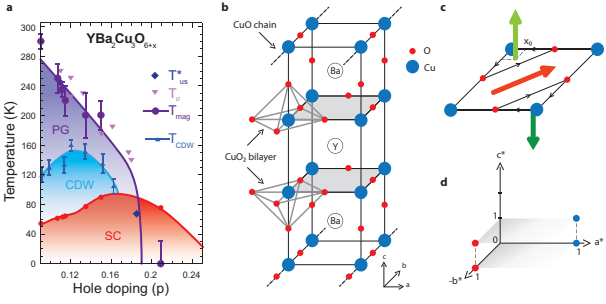}
\caption{
(color online) {\bf $\rm YBa_2Cu_3O_{6+x}$ phase diagram and structure, and the loop currents as a possible model for the cuprates.} (a) $\rm YBa_2Cu_3O_{6+x}$ phase diagram as a function of hole doping (p), showing the pseudo-gap (PG), the incipient charge density wave (CDW) and superconducting (SC) phases. Are reported: $\rm T^{\star}_{us}$ from resonant ultrasound measurements \cite{Shekhter2013}, $\rm T_{mag}$ the temperature of the magnetic IUC order \cite{Baledent2011,Mangin-Thro2015}, $\rm T_{\nu}$ the onset of a-b anisotropy from Nernst effect \cite{Daou2010} and $\rm T_{CDW}$ the onset of CDW correlations from  resonant X-ray measurements \cite{Blanco-Canosa2014}. (b) Crystal structure of the bilayer compound $\rm YBa_2Cu_3O_{6+x}$ with the CuO chains running along ${\bf b}$. (c) Loop current model \cite{Varma2006,Shekhter2009}: each loop induces an orbital magnetic moment $\bf M_i$ (green arrows) perpendicular to the $\rm CuO_2$ plaquette, located at the triangle center, $\rm x_0 = 0.146$. The red arrow represents the associated anapole or toroidal moment $\rm {\bf T} \simeq \sum_{i} {\bf M}_i \times {\bf r}_i$ ($\bf r_i$ stands for the vector connecting the center of the unit cell and the location of the i-th moment). (d) Location of the studied magnetic Bragg reflexions: wave-vectors, given in reduced lattice units, of the form of $\rm {\bf Q}=(1,0,L)$ (blue circles) and $\rm {\bf Q}=(0,1,L)$ (red circles) have been  studied. 
}
\label{aniso-Fig1}
\end{figure}

Upon doping with charge carriers, the lamellar copper oxides evolve from antiferromagnetic (AF) Mott insulators to high temperature superconductors (SC). On the underdoped side of their phase diagram (Fig.~\ref{aniso-Fig1}.a), hole-doped cuprates exhibit unusual electronic and magnetic properties in the so-called pseudo-gap (PG) phase below $\rm T^{\star}$~\cite{revue}. Among cuprate families, various studies in $\rm YBa_2Cu_3O_{6+x}$ (YBCO) have enabled researchers to get a particularly deep understanding of the PG phase. This bilayer system, whose structure is shown in Fig.~\ref{aniso-Fig1}.b, becomes weakly orthorhombic owing to the formation of CuO chains upon increasing oxygen stoichiometry from x=0 to 1, but the $\rm CuO_2$ layers are commonly believed to retain a nearly tetragonal structure which leaves room for spontaneous breaking of the $\rm C_4$ rotational symmetry (into $\rm C_2$) in the electronic and/or magnetic structure. As a strain field, the weak orthorhombicity can facilitate observation of such symmetry breaking by eliminating one of the two possible domains, yielding an a-b anisotropy of physical properties that is much more pronounced than the structural orthorhombicity itself. Such an anisotropy has been reported in electrical transport \cite{Ando2002}, spin dynamics \cite{Hinkov2007,Hinkov2008,Haug2010}, Nernst coefficient \cite{Daou2010,Cyr-Choiniere2015} and nuclear magnetic resonance \cite{Wu2015} measured on detwinned single crystals. In the PG state of another bilayer cuprate $\rm Bi_2Sr_2CaCu_2O_{8+\delta}$ \cite{Lawler2010}, scanning tunneling microscopy also highlighted an intra-unit-cell (IUC) electronic nematic state with unbalanced electronic density on oxygen sites along $\rm {\bf a}$ and $\rm {\bf b}$.

The breaking of time reversal symmetry (TRS) is another feature of the PG physics. Indeed, an IUC magnetic order develops below a temperature $\rm T_{mag}$, matching  $\rm T^{\star}$, as reported by polarized neutron diffraction in four cuprates families \cite{Fauque2006,Mook2008,Li2008,Review2011,Review2013,Baledent2011}. In YBCO, this order is long-ranged at low doping \cite{Mook2008,Review2011}, becomes short-ranged around optimal doping \cite{Mangin-Thro2015} and vanishes at high doping. This IUC magnetic order indicates that translation invariance is preserved, but TRS is broken in the PG state. In addition, resonant ultrasound measurements reported a weak anomaly at $\rm T^{\star}$, indicating that PG phase is a true broken symmetry state \cite{Shekhter2013}. Recently, optical second-harmonic generation measurements in YBCO have further reported a global broken inversion symmetry at $\rm T^{\star}$~\cite{Hsieh2016}, confirming that the pseudogap region coincides with an hidden order. Among other theoretical proposals \cite{Moskvin2012,Lovesey2015,Fechner2016}, the most consistent interpretation of the IUC magnetism \cite{Mangin-Thro2015} remains the loop current (LC) model for the PG state \cite{Varma2006,Shekhter2009,He2012}. It is found to coexist with electronic nematic order  \cite{Fischer2011} as well as charge density waves states \cite{Agterberg2015,Carvalho2016}. The most promising type of LC pattern consists of two counter-propagating LCs flowing over copper and neighboring oxygen sites within each $\rm CuO_2$ unit cell, producing a pair of out-of-plane staggered orbital magnetic moments ($\bf M_i=\pm {\bf M}$) separated along a given diagonal (Fig.\ref{aniso-Fig1}.c). For a single $\rm CuO_2$ layer, four degenerate LC patterns exist, identified by their toroidal moment or anapole \cite{Shekhter2009}, $\rm {\bf T}= \sum_i {\bf r_i} \times {\bf M_i}$ (red arrow in Fig.~\ref{aniso-Fig1}.c) along the other diagonal, along which the inversion symmetry is also broken. The associated IUC magnetic structure factor probed by neutron diffraction can therefore be anisotropic along both diagonals, but no a-b anisotropy is expected, as far as a single $\rm CuO_2$ layer is concerned. 

Motivated by the fact that in underdoped YBCO for a hole doping larger than p$\sim 0.1$, both the a-b anisotropy in Nernst coefficient\cite{Daou2010} and the IUC magnetic order set in below $\rm T^{\star}$ (Fig.\ref{aniso-Fig1}.a), we have carried out a polarized neutron diffraction in a detwinned YBCO sample. We observe an a-b anisotropy in the IUC magnetic structure factor with distinct magnetic intensities along a$^*$ and b$^*$ which show that the mirror symmetry of the CuO$_2$  bilayers is broken below $\rm T^{\star}$. Our data can be described by a crisscrossed arrangement of loop currents within the $\rm CuO_2$ bilayer, with a resulting toroidal axis along the CuO chain, $\rm {\bf b}$, direction.

 %\bigbreak  {\bf Results}

We here report polarized neutron measurements on a low doped $\rm YBa_2Cu_3O_{6.6}$ ($\rm T_c = 63K$,  p=0.12) detwinned single crystal, previously used to study the spin dynamics \cite{Hinkov2007}. The polarized neutron experiments have been performed on the triple axis spectrometer 4F1 (Orph\'ee, CEA-Saclay). A polarizing super-mirror (bender) and a Mezei flipper are inserted on the incoming neutron beam in order to select neutrons with a given spin. In addition, a filter (Pyrolytic Graphite) is put before the bender to remove high harmonics. After the sample, the final polarization, ${\bf P}$, is analyzed by an Heusler analyzer. The incident and final neutron wave vector are set to $\rm k_I = k_F = 2.57\AA^{-1}$. Following previous studies \cite{Fauque2006,Mook2008,Li2008,Review2011,Review2013,Baledent2011,Mangin-Thro2015}, the search for magnetic order in the pseudo-gap phase is performed on Bragg reflections ${\bf Q}$ = (1,0,L)/(0,1,L) with integer L=0,1 values.  The general methods to extract the IUC magnetic signal have already been discussed in 
Refs. \cite{Fauque2006,Mook2008,Li2008,Review2011,Review2013,Baledent2011,Mangin-Thro2015}, the important steps for our analysis are reported in the  Supplemental Material\cite{suppl-mater}.

In order to compare the magnetic signals along the directions $\rm {\bf a}^{\star}$=[1,0] (blue symbols in Fig.~\ref{aniso-Fig1}.d) and $\rm {\bf b}^{\star}$=[0,1] (red symbols), measurements were carried out at Bragg reflections of the form of $\rm {\bf Q} = (1,0,L)$/$\rm (0,1,L)$ with L=0 or 1 in reduced lattice units. Here, we focus on the scattered magnetic intensity for two neutron spin polarizations $\bf P$ (see Supplemental Material\cite{suppl-mater}) (i) $\rm {\bf P} \parallel {\bf Q}$ which measures the full magnetic scattering intensity, (ii) $\rm {\bf P} \perp {\bf Q}$ in the scattering plane, where predominantly the out-of-plane magnetic component, $\rm {\bf M_c}$, is probed. 

\begin{figure}
\includegraphics[width=8cm]{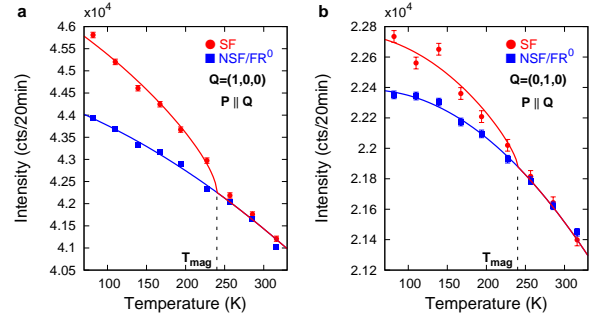}
\caption{
(color online) {\bf Raw Bragg peaks intensity.} Temperature dependence of the Spin-Flip (SF) (red circles) and Non-Spin-Flip (NSF) (blue squares) neutron intensity for a neutron polarization $\rm {\bf P} \parallel {\bf Q}$. a) At $\rm {\bf Q} = (1,0,0)$ (along $\rm {\bf a}^{\star}$). b) At $\rm {\bf Q} = (0,1,0)$ (along $\rm {\bf b}^{\star}$). In the SF channel, a magnetic signal is observed below $\rm T_{mag} \sim 240 K$ on top of the NSF intensity normalized at high temperature by a constant flipping ratio, FR$^0 \sim 40$. Data have been averaged over a temperature range of 25K to improve the statistics. Error bars of standard deviation are about the size of the points.
}
\label{aniso-Fig2}
\end{figure}

Fig.~\ref{aniso-Fig2} shows the raw neutron intensity on two Bragg peaks along the directions $\rm {\bf a}^{\star}$ and  $\rm {\bf b}^{\star}$ for L=0 and $\rm {\bf P} \parallel {\bf Q}$. The intensities for 2 neutron spin states are shown, in the spin flip (SF) channel when the scattering process flips the neutron spin at the sample position and non spin flip (NSF) when it is preserved. Following a standard procedure~\cite{Fauque2006,Mook2008,Li2008,Review2011,Review2013,Baledent2011,Mangin-Thro2015}, both curves have been normalized at high temperature over some temperature range (here between 250K and 330K). On the one hand, NSF measures the nuclear Bragg peak intensity which exhibits a continuous decay when increasing the temperature as expected for a Debye-Waller factor. As the sample is detwinned, the NSF intensity along $\rm {\bf b}^{\star}$ (Fig.~\ref{aniso-Fig2}.b) is weaker than along  $\rm {\bf a}^{\star}$ (Fig.~\ref{aniso-Fig2}.a)). On the other hand, the SF scattering intensity probes a true SF magnetic scattering (if any) on top of a polarization leakage of the NSF channel into the SF channel. The latter is given by the NSF intensity divided by the flipping ratio FR$^0$. For a perfectly spin polarized neutron beam, FR$^0$ goes to infinity and the leakage vanishes. On top on the normalized nuclear scattering, the SF intensity then exhibits an extra scattering at low temperature that is attributed to the IUC magnetic component (Fig.~\ref{aniso-Fig2}). In both $\rm {\bf a}^{\star}$ and $\rm {\bf b}^{\star}$ directions, the magnetic signal appears below $\rm T_{mag} \sim 240K$, in agreement with $\rm T_{mag}=220K \pm 20K$ determined in an early study on the same sample matching T* 
$\sim$ 230K deduced from resistivity measurement at that doping \cite{Fauque2006}. Qualitatively, the magnetic signal for $\rm {\bf Q} = (0,1,0)$  is much weaker than  for $\rm {\bf Q} = (1,0,0)$, which underlines an a-b anisotropy of the Q=0 IUC magnetic signal. 

\begin{figure}
\includegraphics[width=8cm]{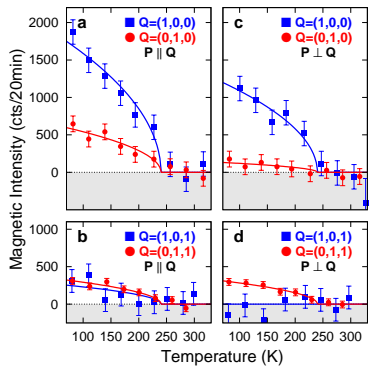}
\caption{
(color online) {\bf Temperature dependence of the IUC magnetic intensity in both directions $\rm {\bf a}^{\star}$ (blue squares) and $\rm {\bf b}^{\star}$ (red circles)}. (left panels) polarization $\rm {\bf P} \parallel {\bf Q}$: (a) at L=0 and (b) at L=1. (right panels) polarization $\rm {\bf P} \perp {\bf Q}$: (c) at L=0 and (d) L=1. Magnetic intensities are obtained using the procedure described in  Supplemental Material\cite{suppl-mater}. Each curve is described by the function $\rm I_0 (1 - T/T_{mag})^{2\beta}$ where $I_0$ is fitted with $\rm \beta = 0.25$ and $\rm T_{mag} = 240K$ being fixed. Below this temperature, the amplitude of the magnetic intensity $\rm I_{mag}$ differs as a function of wave-vector and polarization. Error bars are of standard deviation. 
}
\label{aniso-Fig3}
\end{figure}

To perform a more quantitative analysis of the magnetic scattering, we need to calibrate the change of neutron polarization with temperature. We perform a systematic analysis based on generic procedure improved in previous studies~\cite{Baledent2011,Mangin-Thro2015}. We then determined the IUC magnetic intensity, that we report in Fig.~\ref{aniso-Fig3}, at 4 different Bragg spots for 2 different neutron polarization states. Fig.~\ref{aniso-Fig3}.a shows the full magnetic intensity for $\rm {\bf P} \parallel {\bf Q}$ at L=0 as a function of temperature. Quite remarkably, the full magnetic intensity exhibits a net difference between the two directions, being $\rm \sim 3$ times larger at $\rm {\bf Q} = (1,0,0)$ than at $\rm (0,1,0)$. Increasing L to 1, Fig.~\ref{aniso-Fig3}.b, the magnetic intensity becomes almost identical in both directions, with slightly more intensity for $\rm (0,1,1)$. A net a-b anisotropy of the IUC magnetic intensity thus exists, but, remarkably, changes as a function of L.

Rotating  $\rm {\bf P} \perp {\bf Q}$ in the scattering plane, one selectively probes the scattering intensity $\rm \propto {\bf M}_{c}^2$, which corresponds to the out-of-plane components of the magnetic moments. In Fig.~\ref{aniso-Fig3}.c, the magnetic intensity is at least 6 times larger at $\rm {\bf Q} = (1,0,0)$ than at $\rm  (0,1,0)$, where the magnetic intensity vanishes within error bars. At L=1 (Fig.~\ref{aniso-Fig3}.d), the a-b anisotropy is fully reversed. Therefore, the anisotropy is more pronounced for the polarization $\rm {\bf P} \perp {\bf Q}$ than for the polarization $\rm {\bf P} \parallel {\bf Q}$. Clearly, the out-of-plane components of the magnetic moments are mainly responsible for the observed a-b anisotropy varying with  L (Results on the in-plane moment, which exhibits less anisotropy, will be presented elsewhere). 
% \bigbreak {\bf Discussion}

\begin{figure}
\includegraphics[width=8cm]{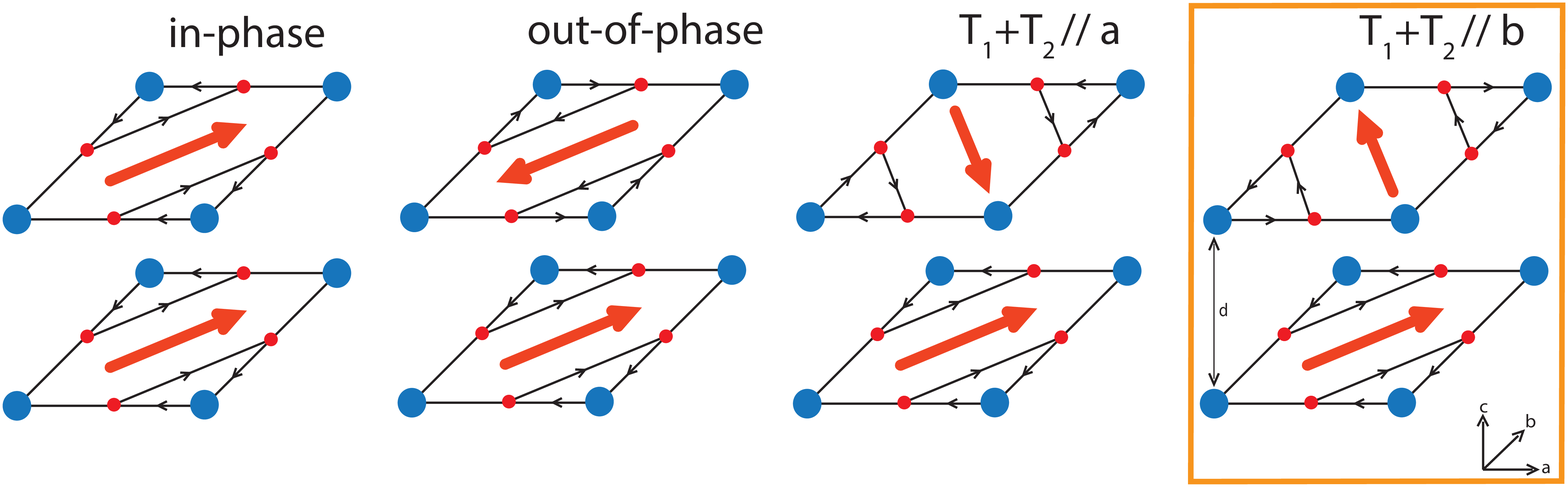}
\caption{
(color online) {\bf Possible model for the a-b anisotropy of the IUC magnetic intensity.} Choosing one possible loop current state over the four existing ones for the lower plane (1) and varying the state for the upper plane (2), one obtains four different types of configurations: in-phase, out-of-phase anapoles/LC correlations, and two situations where the sum of anapoles in both planes gives a resultant either along $\rm {\bf a}$ or $\rm {\bf b}$. The last one is selected by the experimental results. Each of these configurations are 4 times degenerate as one could select a different LC state for the lower plane. However, the upper plane configuration would be always correlated with the first plane as discussed here. The LC structure factor depends only on the relative configurations of both planes (see  Supplemental Material\cite{suppl-mater}). }
\label{aniso-Fig4}
\end{figure}

\begin{table}
\begin{tabular}{|c|c|c|c|c|}
\hline
LCs& in-phase & out-of-phase & $\rm ({\bf T_1 + T_2}) \! \! \parallel \! \! {\bf a}$ & $\rm ({\bf T_1 + T_2}) \! \! \parallel \! \! {\bf b}$\\
\hline
$\rm (1,0,L)$ & $\rm \propto {\bf cos^2}(\pi \frac{d}{c} L)$ & $\rm \propto {\bf sin^2}(\pi \frac{d}{c} L)$ & $\rm \propto {\bf sin^2}(\pi \frac{d}{c} L)$ & $\rm \propto {\bf cos^2}(\pi \frac{d}{c} L)$\\
\hline
$\rm (0,1,L)$ & $\rm \propto {\bf cos^2}(\pi \frac{d}{c} L)$ & $\rm \propto  {\bf sin^2}(\pi \frac{d}{c} L)$ & $\rm \propto {\bf cos^2}(\pi \frac{d}{c} L)$ & $\rm \propto {\bf sin^2}(\pi \frac{d}{c} L)$\\
\hline
\end{tabular}
\caption{L-dependence of the out-of-plane magnetic structure factor for each group of LC magnetic patterns (different bilayer correlation) of Fig. \ref{aniso-Fig4} at $\rm {\bf Q}=(1,0,L)$ and $\rm {\bf Q}=(0,1,L)$ (see the structure factor calculation in Supplemental Material\cite{suppl-mater}).}
\label{structure}
\end{table}

Our study puts stringent constraints on the possible nature of the IUC magnetic order: (i) the magnetic structure factor has to be maximum at L=0, (ii) the IUC order must produce a scattering pattern characterized by a remarkable a-b anisotropy varying with L, as revealed by a closed comparison of magnetic intensities at (1,0,L) and (0,1,L). The change of the structure factor along  $\rm {\bf c}^{\star}$ - odd in L along  $\rm {\bf a}^{\star}$  and even in L along $\rm {\bf b}^{\star}$ -suggests that the bilayer structure of IUC order does not represent the symmetries of the higher-temperature phase {\it i.e.} that the symmetry is truly broken. More specifically, that indicates that the bilayer mirror plane across the Y site is broken in the pseudogap phase of YBCO. In particular, the abovementionned requirements cannot be fulfilled by magnetic nematic states, that involve spin or orbital moments located on oxygen sites \cite{Fauque2006,Moskvin2012}. Indeed, these magnetic patterns fail to reproduce a L-dependent a-b anisotropy for the out-of-plane magnetic scattering intensity. This is at variance with a crisscrossed arrangement of loop currents that we describe below. 

The LC model naturally induces staggered out-of-plane magnetic moments. For a single $\rm CuO_2$ layer, as discussed above, there are 4 LC degenerate patterns, none of which is expected to give rise to the observed a-b anisotropy. However, in a bilayer system such as YBCO,  one needs to consider the relative arrangement of LC patterns in the two $\rm CuO_2$ layers, labeled (1) and (2), separated along the $\rm {\bf c}$ axis by a distance d=0.28c. This yields to 4x4 possible magnetic configurations, which can be classified into four distinct groups of LC patterns (Fig.~\ref{aniso-Fig4}) and identified by the resulting toroidal axis $\rm ({\bf T}_1 + \bf {T}_2)$. Each group breaks different symmetries: in the first two groups, the different configurations are connected by $\rm C_4$ rotation, whereas in the last two groups the configurations are connected by $\rm C_2$ rotation (equivalent to time-reversal) and breaks the bilayer mirror symmetry. Each group is characterized by a specific magnetic structure factor shown in Table~\ref{structure}. 

If within the bilayer LC patterns are either in-phase or out-of-phase (where $\rm {\bf T_1}$  and $\rm {\bf T_2}$ are parallel or anti-parallel), the scattered magnetic intensity is respectively modulated by $\rm |2 cos(\pi \frac{d}{c} L)|^2$ or $\rm |2 sin(\pi \frac{d}{c} L)|^2$ along both $\rm {\bf a}^{\star}$ or $\rm {\bf b}^{\star}$. The in-phase configuration has been previously favored \cite{Fauque2006,Review2011,Review2013,Baledent2011,Mangin-Thro2015} since the IUC intensity is the strongest at Bragg peaks with L=0 (that actually dismisses the out-of-phase case). However, we here observe that the IUC intensity varies with L differently along the directions $\rm {\bf a}^{\star}$ and $\rm {\bf b}^{\star}$. Both of these configuration groups are thus inconsistent with our observations. 

The two other configuration groups are featured by a crisscrossed arrangement of LC patterns, and are further identified owing to the orientation of their resulting toroidal axis $\rm ({\bf T_1}$ + $\rm {\bf T_2})$ parallel to either $\rm {\bf a}$ or $\rm {\bf b}$. These two configurations are breaking the mirror symmetry at the middle of the bilayers (the Y site) as requested by the experiment. Importantly, the energy difference between these two groups is expected to be linearly coupled to the orthorhombicity. Therefore, any orthorhombic distortion would remove the degeneracy between both configurations. Interestingly, they exhibit out-of-phase modulations of the scattering intensities at (1,0,L) and (0,1,L), yielding a L-dependent a-b anisotropy of the scattering intensity (Tab.~\ref{structure}). Only the configuration $\rm ({\bf T_1}$ + $\rm {\bf T_2}) \parallel {\bf b}$ gives rise to a magnetic intensity larger for (1,0,0) than (0,1,0). Using the structure factor of Tab.~\ref{structure}, one remarks that the anisotropy weakens and is even reversed at L=1. This evolution of the magnetic intensities is consistent with the experimental observations (Fig.~\ref{aniso-Fig3}). Nevertheless, a weaker or even null intensity for (1,0,1) is observed (Fig.~\ref{aniso-Fig3}.c). That could simply be due to limited statistics or more interestingly could be related by a larger distance $d/c$ separating the moments. Note that, in models that include the apical oxygens~\cite{Li2008,Weber2009}, the LCs could be delocalized and a larger distance $d/c$ would occur and can further reduce intensity at (1,0,1) than at (0,1,1). Globally, the LC configuration in Fig.~\ref{aniso-Fig4} with $\rm ({\bf T_1}$ + $\rm {\bf T_2}) \parallel {\bf b}$ accounts for the polarized neutron data. This particular arrangement of LC within the bilayer is a truncated version of the chiral LC order considered by Pershoguba {\it et al} \cite{Pershoguba2013} as a pattern of a cholesteric arrangement of toroidal moments. In constrast to our results, the proposed full chiral LC order \cite{Pershoguba2013}  would double the magnetic unit cell along $\bf c$ and shift the magnetic Bragg reflection at half-integer L values. This is not what is happening, instead we here demonstrate that the direction of the resulting toroidal axis $\rm ({\bf T_1}$ + $\rm {\bf T_2})$ is always pinned down along $\rm {\bf b}$, i.e the CuO chain direction. 

On general grounds, it is interesting to relate our finding to the electronic nematicity which has been abundantly discussed in the context of cuprates \cite{Ando2002,Hinkov2007,Hinkov2008,Haug2010,Daou2010,Cyr-Choiniere2015,Wu2015} even occuring within the unit cell\cite{Lawler2010}. Our study shows that the resulting toroidal moment  for a bilayer  presents an Ising anisotropy along the CuO chain. By itself, the presence of an a-b anisotropy that pins the direction of a vector magnetic order parameter does not amount to nematicity because nematic order is characterized by a director, and not a vector order parameter, such as the anapoles of Fig. \ref{aniso-Fig4}. However, we show that the magnetic structure of the IUC order reveals an unexpected a-b anisotropy due to the bilayer structure of YBCO and its weak orthorhombic distortion. Therefore, our study suggests that there could exist an interplay between a crisscrossed LC order and the reported nematicity in YBCO\cite{Ando2002,Hinkov2007,Daou2010,Cyr-Choiniere2015}, an intriguing scenario that has not been consider so far. One can further notice the good agreement of the nematicity deduced from Nernst effect with the onset of the IUC order (Fig.~\ref{aniso-Fig1}.a) that explicitly occur at higher temperature than the charge density wave (CDW) signal \cite{Blanco-Canosa2014}. Using a single band model to discuss instabilities of a weakly correlated Fermi liquid~\cite{Oganesyan2001}, the nematicity appears as a spontenaous distortion of the Fermi surface (d-wave Pomeranchuck instability). Within the 3-band Emery model,  it has been shown that  intra-unit-cell instabilities such as LC order and electronic nematicity could coexist~\cite{Fischer2011}. It might be interesting to re-examine the role of the bilayer in light of our results. 

Finally, the charge order  that develops well below T* in YBCO also exhibits an a-b anisotropy\cite{Wu2015} with a remarkable L-dependent a-b anisotropy of its superstructure reflections in hard-X ray diffraction measurements~\cite{Forgan2015}. The analysis and modeling of X-ray diffraction measurements \cite{Forgan2015} actually indicate as well that the reported quasi-2D CDW breaks the mirror symmetry of the CuO$_2$ bilayers in YBCO as does the crisscrossed LC states that we are reporting here. This rises questions concerning a possible interplay between CDW  and the crisscrossed LC states.

\begin{acknowledgments}

We thank Marc-Henri Julien, Bernhard Keimer, Arkady Shekhter, Chandra Varma and Victor Yakovenko for valuable discussions. We also acknowledge financial support from grants of the Agence Nationale pour la Recherche (ANR): UNESCOS (contract ANR-14-CE05-0007) and  NirvAna (contract ANR-14-OHRI-0010). 

\end{acknowledgments}

$^*$ current address: Institut Laue-Langevin, 71 avenue des martyrs, 38000 Grenoble, France\\
$^\#$ current address: International Center for Quantum Materials, School of Physics, Peking University, Beijing 100871, China

\clearpage

\centerline{\label{4F1}\bf Supplemental Materials:}

We here give additional information concerning the sample and the polarized neutron scattering setup used during the experiment. The latter aspect has  been already widely discussed in Refs. \cite{Fauque2006b,Mook2008b,Baledent2011b,Li2008b,Li2011b,DeAlmeida-Didry2012b,Review2011b,Review2013b,Mangin-Thro2014b,Mangin-Thro2015b} in the context of the intra-unit-cell magnetic order in cuprate high temperature superconductors.

\section*{\bf A. Sample}

The sample is made of an array of 180 individually detwinned $\rm YBa_2Cu_3O_{6.6}$ single crystals, co-aligned and glued on an aluminum plate. The sample is characterized by a volume of $\rm \sim 450 mm^{3}$, with a mosaic of $\rm 2.2^{\circ}$. Its superconducting critical temperature, $\rm T_c$, determined by neutron depolarization technique, is 63 K, corresponding to a hole doping level of $p \simeq 0.12$. The same sample was used to first evidence the IUC magnetic order \cite{Fauque2006b}, and also to study the a-b anisotropy of the spin dynamics around the antiferromagnetic wave-vector in both superconducting and normal state \cite{Hinkov2007b}. The  twin-domain population ratio was 94:6. \cite{Hinkov2007b}. Between those previous studies and the present one, the array of single crystals was re-aligned, yielding an increase of the sample mosaic from $\rm 1.2^{\circ}$ to $\rm 2.2^{\circ}$. This leads to a weaker Bragg peak intensities and a relative increase of the background. However, the intrinsic physical properties of the twin-free single crystal are preserved.

\section*{\bf B. Polarized neutron scattering setup on 4F1}

The polarized neutron scattering measurements were performed on the cold neutron triple-axis spectrometer 4F1 at Orph\'ee reactor (CEA-Saclay, France). The spectrometer is equipped with a double Pyrolitic Graphite (PG002) monochromator, followed by a polarizing mirror (bender) to polarize the incident neutron spins and a polarizing Heusler analyzer is used to probe the spin state of scattered neutrons. The incident and final neutron wave-vectors are set to $\rm k_I = k_F = 2.57\AA^{-1}$ and the energy resolution is $\sim 1meV$. Before the bender, a PG filter removes higher harmonics from the incident neutron beam. After the bender, a Mezei flipper is used to flip the neutron spins. All along the neutron path, the neutron spin polarization is controlled by magnetic guide fields of a few Gauss. Around the sample, standard XYZ Helmholtz-like coils control the neutron spin polarization. The sample is attached on the cold head of a closed-circle cryostat.

Intensities were measured at wave-vectors of the form $\rm {\bf Q}$ = (H,0,L) and $\rm {\bf Q}$ = (0,K,L) (with L=0 and 1). We did not investigate the Bragg peaks for L=2 as the Q=0 magnetic signal at that position is negligible~\cite{Mook2008b,Review2011b}. The scattering wave-vector $\rm {\bf Q}$=(H,K,L) is given in reduced lattice units ($\rm \frac{2 \pi}{a}, \frac{2 \pi}{b}, \frac{2 \pi}{c} $), with $\rm a=3.82$\AA, $\rm b=3.87$\AA and $\rm c=11.7$\AA. The scattering vector $\rm {\bf Q}$ is here a wave vector of the reciprocal lattice. In the present study, we are dealing with an IUC magnetic order, i.e a $\rm {\bf Q}$=0 antiferromagnetic magnetic order. This needs to be specified in order to avoid any confusion with the usual antiferromagnetic spin correlations located in moment space at the planar wave vector $\rm {\bf q}_{AF}$=(0.5,0.5)$\equiv (\pi,\pi)$. 

The sample has been aligned in two different scattering planes in order to reach the different wave-vectors. First, the scattering plane [1,0,0]/[0,1,0], with [0,0,1] direction perpendicular to the scattering plane, allows to access $\rm {\bf Q}$ = (H,0,0) and $\rm {\bf Q}$ = (0,K,0). Second, for the study of $\rm {\bf Q}$=(H,0,1) and (0,K,1), the sample spans in the [-1,1,0]/[1,1,2] scattering plane, with the direction [-$\epsilon$,-$\epsilon$,1] perpendicular to the scattering plane ($\rm \epsilon = (\frac{a}{c})^2 \sim 0.11$). 

%\newpage \section*{\label{polar} Supplementary Note 2}

\section*{\bf C. Polarization analysis}

Polarized neutron scattering is a powerful technique to study magnetic structures owing to the interaction of the neutron spin with static or dynamic magnetic field distribution present within a sample~\cite{squiresb}. The scattered magnetic intensity provides a measurement of the Fourier transform of the magnetic correlation function. Considering a distribution of magnetic moments, only the correlation function of their magnetic component perpendicular to $\rm {\bf Q}$ can be probed, owing to the dipolar nature of the scattering potential. The scattered magnetic intensity is thus proportional to $\rm | {\bf \sigma}.{\bf M}_{\perp}(Q) | ^2$~\cite{squiresb}, where $ {\bf M_{\perp}(Q)} = {\bf {\hat Q}} \times  {\bf M(Q)} \times {\bf {\hat Q}}$ stands for the Fourier transform of the magnetic moment distribution perpendicular to the unitary vector ${\bf {\hat Q}}$=(H,K,L)/$|$Q$|$. The neutron spin is described in terms of Pauli matrices ${\bf \sigma}$ and the polarization $\rm {\bf P}$ specifies the direction of the quantification axis. As a consequence, only the component of $ {\bf M_{\perp}(Q)}$ perpendicular to the neutron spin polarization vector $\rm {\bf P}$ is spin flip (SF), whereas the component which is parallel is non spin flip (NSF). At variance with the magnetic scattering, the nuclear scattering does not affect the neutron spin and is therefore always in the NSF channel. Using a standard XYZ polarization analysis, one sets $\rm {\bf P}$ in three orthogonal orientations (X,Y,Z), where  $\rm {\bf P}$ is respectively parallel to $\rm {\bf {\hat Q}}$ (X), perpendicular to $\rm {\bf {\hat Q}}$ but still in the scattering plane (Y) (that is the magnetic component $\perp$ $\rm {\bf {\hat Q}}$ reported in the Fig. 3.c-d of the manuscript), and parallel to the vertical direction (Z). For the X polarization, the full magnetic scattering is SF. The Y and Z polarization probes in the SF channel two complementary scattered magnetic intensity, whose sum is equal to the full scattered magnetic intensity. This is often referred to as the polarization sum rule, which is valid in the absence of chirality only~\cite{Review2011b,squiresb}.

Considering the case of an ordered state with magnetic moments $\rm {\bf M}=({\bf M}_a,{\bf M}_b,{\bf M}_c)$, the magnetic intensity measured in the SF channel with the polarization X is the total measured magnetic intensity~\cite{Review2011b,squiresb}:

\begin{widetext}
\begin{equation}
I_{mag,X} \propto |F_M({\bf Q}) f(Q)|^2 \Bigg \{ \left[1-\left(\frac{2 \pi}{a}\frac{H}{|{\bf Q}|}\right)^2\right] {\bf M}_a^2 + \left[1-\left(\frac{2 \pi}{b}\frac{K}{|{\bf Q}|}\right)^2\right] {\bf M}_b^2 + \left[1-\left(\frac{2 \pi}{c}\frac{L}{|{\bf Q}|}\right)^2\right] {\bf M}_c^2 \Bigg \}
\end{equation}
\end{widetext}

$F_M({\bf Q})$ is the magnetic structure factor. Below in section E,  that structure factor is calculated in the loop current model for a CuO$_2$ bilayer. $f(Q)$ is the magnetic form factor vanishing at large $|Q|$. It corresponds to the Fourier transform of the local magnetic moment extension in real space. The terms between brackets are the neutron orientation factor because neutrons are scattered by only magnetic component perpendicular to $\rm {\bf {\hat Q}}$. For the polarizations Y and Z where $\rm {\bf P \perp {\hat Q}}$, a fraction of the magnetic intensity is transferred from the SF to the NSF channel. Thus, for a given polarization direction $\rm \alpha=(X,Y,Z)$,

\begin{equation}
I_{mag,\alpha} \propto A^{\alpha} {\bf M}_a^2 + B^{\alpha} {\bf M}_b^2 + C^{\alpha} {\bf M}_c^2
\end{equation}

The coefficients $\rm A^{\alpha}$, $\rm B^{\alpha}$ and $\rm C^{\alpha}$ depend on the polarization direction and on the transferred momentum through the orientation factor. They are calculated for $\rm {\bf Q}$ = (1,0,L) and $\rm {\bf Q}$ = (0,1,L), and gathered in the following table (Tab.~\ref{coeff}). These coefficients depend on the chosen horizontal scattering plane and the corresponding vertical direction. As discussed above in section B., the direction perpendicular to the scattering plane was (0,0,1) for the studies with L=0 and [-$\epsilon$,-$\epsilon$,1] (with $\rm \epsilon \sim 0.11$) for the studies with L=1. 

\begin{table}[h]
\renewcommand*{\thetable}{S\Roman{table}}
\begin{tabular}{|c||c||c|c|c||c||c|c|c|}
\hline
$\alpha$ & ${\bf Q}$ & A & B & C & ${\bf Q}$ & A & B & C\\
\hline
\hline
X & (1,0,0) & 0 & 1 & 1 & (0,1,0) & 1 & 0 & 1\\
\hline
Y & (1,0,0) & 0 & 0 & 1 & (0,1,0) & 0 & 0 & 1\\
\hline
Z & (1,0,0) & 0 & 1 & 0 & (0,1,0) & 1 & 0 & 0\\
\hline
\hline
X & (1,0,1) & 0.1 & 1 & 0.9 & (0,1,1) & 1 & 0.1 & 0.9\\
\hline
Y & (1,0,1) & 0.01 & 0.09 & 0.82 & (0,1,1) & 0.09 & 0.01 & 0.82\\
\hline
Z & (1,0,1) & 0.09 & 0.91 & 0.08 & (0,1,1) & 0.911 & 0.09 & 0.02\\
\hline
\end{tabular}
\caption{Weight of the different magnetic correlation functions as a function of the neutron spin polarization and the transfered momentum ${\bf Q}$.}
\label{coeff}
\end{table}

It appears for both $\rm {\bf Q}$ = (1,0,L) and = (0,1,L), that the polarization direction Y essentially corresponds to the magnetic moment component $\rm {\bf M}_c^2$. This is the quantity reported in Fig. 3c-d. For L=0, Y even always probe $\rm {\bf M}_c^2$ whereas Z probes the in-plane component ($\rm {\bf M}_a^2$ for (0,1,0) and $\rm {\bf M}_b^2$ for (1,0,0)). Whatever the transfered momentum, the sum of the scattered magnetic intensity for Y and Z polarization correspond to the full scattered magnetic intensity given by the X polarization (given by Eq. 1 above) (polarization sum rule)~\cite{Review2011b,squiresb}.
		
%\section*{\label{analysis} Supplementary Note 3}
\section*{\bf D. Data analysis}

We here briefly give the method to extract the magnetic signal, $\rm I_{mag}$. That analysis follows the procedure which has been established to sucessfully evidence the intra-unit-cell (IUC) magnetic order over a wide range of hole doping levels and cuprate families~\cite{Fauque2006b,Mook2008b,Baledent2011b,Li2008b,Li2011b,DeAlmeida-Didry2012b,Review2011b,Review2013b,Mangin-Thro2014b,Mangin-Thro2015b}. Here, it is only assumed that there is no magnetic scattering at high temperature (here between $\rm T_{mag} \sim 240K$ and 330K). The same hypothesis is actually employed for the determination of the pseudogap temperature, T*, in various physical properties like in resistivity measurements or for the uniform magnetic susceptibilty measured by Knight shift in Nuclear Magnetic Resonance experiments~\cite{Review2011b}.

While for a perfectly polarized neutron beam, only a magnetic scattering can show up in the SF channel, in a real polarized neutron scattering measurement, the polarization of the neutron beam is not perfect but finite $p\sim 0.96$. That defines a flipping ratio $\rm FR=I^{NSF}/I^{SF}=(1+p)/(1-p)$. For our setup on 4F1, a typical flipping ratio is of the order of 40-50. As a consequence, the nuclear Bragg intensity in the NSF channel, $\rm I^{NSF}$, leaks into the SF channel for measurements at Bragg positions. It is then convenient to report at each temperature the quantity 1/FR(T) which is the ratio of measured scattered SF and NSF intensities. For each polarization direction $\rm \alpha=(X,Y,Z)$, that reads: 
\begin{equation}
1/{FR_{\alpha}(T)}= I^{SF}_{\alpha}/I^{NSF}_{\alpha} = I_{mag,\alpha}/{I^{NSF}_{\alpha}}+ 1/{FR^0_{\alpha}}
\label{mag1}	
\end{equation}

\begin{figure}[h]
\renewcommand*{\thefigure}{S\arabic{figure}}
\includegraphics[width=9cm]{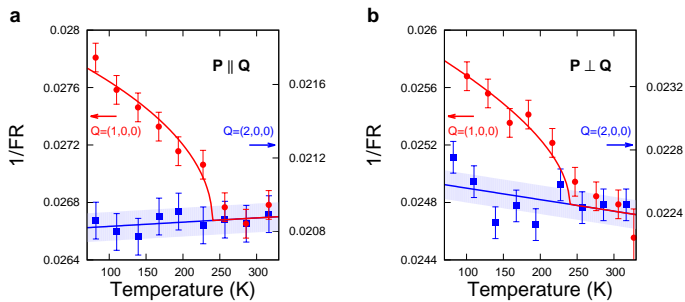}
\caption{
Temperature dependencies of the inverse of the flipping ratio $\rm 1/FR$ (see text) measured at 
$\rm {\bf Q}$ = (1,0,0) (red circles) and at $\rm {\bf Q}$ = (2,0,0) (blue squares). The right vertical scale corresponding to   $\rm {\bf Q}$ = (2,0,0) data has been shifted so that both curves coincide at high temperature (above 250 K).  $\rm 1/FR$ is represented for 2 polarizations: (a) polarization X ($\rm {\bf {P}}$  parallel to $\rm {\bf {\hat Q}}$) (b) polarization Y ($\rm {\bf {P}}$  perpendicular to $\rm {\bf {\hat Q}}$).  Data are averaged over a temperature range of 25K to improve the statistics. 
}
\label{aniso-sup-Fig1}
\end{figure}

$\rm 1/FR^0_{\alpha}$ represents the quality of the polarization of the neutron beam. Experimentally, it turns out that this quantity varies slightly for the different neutron spin polarization. It can as well exhibits a weak temperature dependence (see a detailed discussion in Ref. \cite{Mangin-Thro2014b}). Fig.~\ref{aniso-sup-Fig1} reports $\rm 1/FR_{\alpha}(T)$ measured for 2 Bragg peaks $\rm {\bf Q}$=(1,0,0)  (in red) and $\rm {\bf Q}$=(2,0,0) (in blue), for 2 neutron polarizations (X for  Fig.~\ref{aniso-sup-Fig1}.a and Y for  Fig.~\ref{aniso-sup-Fig1}.b). 
For $\rm {\bf Q}$=(1,0,0), the reported quantity is the ratio of the raw intensity shown in Fig. 2a of the manuscript. The data for $\rm {\bf Q}$=(2,0,0) have been shifted to scale at high temperature with the data at $\rm {\bf Q}$=(1,0,0). We stress again that this is the central assumption of the data analysis that no magnetic scattering exists at high temperature (say here above 250K). 

The behaviour of $\rm 1/FR_{\alpha}(T)$  is clearly different for both Bragg peaks. A weakly linear dependence is observed for $\rm {\bf Q}$=(2,0,0) whereas an additional upturn sets in for $\rm {\bf Q}$=(1,0,0). That signs the existence of a magnetic IUC scattering at that position which (if any) is negligible at large wave vector, $\rm {\bf Q}$=(2,0,0), due to the magnetic form factor~\cite{Review2011b}. The behaviour at $\rm {\bf Q}$=(2,0,0) then defines the evolution of the bare inverse of the flipping ratio, $\rm 1/{FR^0_{\alpha}}$. 

At high temperature, when there is no magnetic intensity, $\rm 1/FR$ equals $\rm 1/FR^0$. Below $\rm T_{mag} \sim 240K$, the magnetic intensity $\rm I_{mag,\alpha}$ appears. Such a value of $\rm T_{mag}$ agrees with $\rm 220K \pm 20K$ determined by previous studies of the IUC magnetic order \cite{Fauque2006b} on the same detwinned sample. The extra magnetic signal displays a characteristic T-dependence that can be described by $\rm I_0 (1 - T/T_{mag})^{2\beta}$\cite{Mook2008b} where $I_0$ is fitted and $\rm \beta = 0.25$ is fixed. 

Once the behaviour of $\rm 1/{FR^0_{\alpha}}$ is known from $\rm {\bf Q}$=(2,0,0) and multiplying both sides of Eq. \ref{mag1} by $\rm I_{\alpha}^{NSF}$, one can determine, $\rm I_{mag,\alpha}$, the magnetic intensity for each polarization. We report  in Fig. 2 of the manuscript the comparison for L=0 between $\rm I_{\alpha}^{SF}$ and $\rm I_{\alpha}^{NSF}/FR_{\alpha}^0$ whose difference directly gives $\rm I_{mag,\alpha}$. This analysis procedure was systematically applied to extract the magnetic scattering at $\rm {\bf Q}$=(1,0,L) and $\rm {\bf Q}$=(0,1,L) for L=0 and 1. The deduced magnetic intensities are shown in Fig. 3 of the manuscript. 

As we here discuss the anisotropy of the magnetic signal, one needs not to confuse with the anisotropy of the nuclear intensities. 
$\rm YBa_2Cu_3O_{6.6}$ exhibits an orthorhombic structure whereas the parent compound $\rm YBa_2Cu_3O_{6}$ has a tetragonal structure.  The additional oxygen atoms (0.6 per formula unit) are forming the CuO chain along  ${\bf b}$. These extra oxygen atoms in the CuO chain modify the nuclear structure factor by adding a term 
${\rm \propto 2 b_O \cos( \pi K)}$, where $\rm b_O$ stands for the oxygen neutron scattering length. This term gives a negative contribution to the structure factor for odd K and a positive one for even K. This is the origin of the a-b anistropy for nuclear Bragg reflection (1,0,L) et (0,1,L), while there is no anisotropy for Bragg reflections (2,0,0) and (0,2,0). As a result, the calculated nuclear intensities are always larger along  $\rm {\bf a}^{\star}$=[1,0] than  $\rm {\bf b}^{\star}$=[0,1]. The measured intensities in the NSF channel are given in the following table (Tab.~\ref{NSFaniso}) for the four wave-vectors ${\bf Q}$ =(1,0,0), ${\bf Q}$=(0,1,0), ${\bf Q}$ =(1,0,1) and ${\bf Q}$ =(0,1,1). For both values of L, the intensities along  $\rm {\bf a}^{\star}$ is larger than along  $\rm {\bf b}^{\star}$. In contrast, the magnetic intensities are not systematically larger along  $\rm {\bf a}^{\star}$.

\begin{table}[h]
\renewcommand*{\thetable}{S\Roman{table}}
\begin{tabular}{|c||c|c||c|c|}
\hline
${\bf Q}$ & (1,0,0) & (0,1,0) & (1,0,1) & (0,1,1)\\
\hline
$I^{NSF}$ & $1.6\times10^6$ & $1.1\times10^6$ & $1.5\times10^6$ & $1.5\times10^5$\\
\hline
\end{tabular}
\caption{NSF intensities measured at ${\bf Q}$ = (1,0,0)/${\bf Q}$ = (0,1,0) and ${\bf Q}$ = (1,0,1)/${\bf Q}$ = (0,1,1).
All numbers are counts/20 minutes. }
\label{NSFaniso}
\end{table}

%  \section*{\label{structurefactor} Supplementary Note 4}
\section*{\bf E. Structure factor associated with the loop currents model in a bilayer system}

Among several possible models, we focus on the loop currents (phase CC-$\Theta_2$ \cite{Varma2006b}) for which the out-of-plane moment is directly related by an anapole or toroidal moment, within each $\rm CuO_2$ plaquette (Fig.1.c). The anapole ${\bf T}$  points along the diagonal within the plane and defines the loop current state in any given CuO$_2$ layer. There are four different loop current states, which are obtained by $\pi/2$-rotation \cite{He2012b} of $\rm {\bf T}$. These loops produce orbital magnetic moment at ($\pm x_0$,$\pm x_0$) or ($\pm x_0$,$\mp x_0$) within the CuO$_2$ plane (Fig.~1.c). These four equivalent configurations exhibit indistinguishable structure factors at Bragg positions (1,0,L) or (0,1,L) where the IUC magnetic order has been observed \cite{Review2011b}. 

Further, the YBCO structure has a bilayer of CuO$_2$ planes (Fig.~1.a). The two CuO$_2$ planes within the unit cell of the YBCO compound are $\pm \frac{d}{2}$ away from the yttrium site (d=0.28c=3.3$\AA$). As it exists four possible states for each plane, it then yields to 16 possible configurations of the LCs for a bilayer system. Whatever is the LC state in a given plane, one can determine 4 different groups corresponding to the type of correlations of the LC or toroidal moment from one layer ${\bf T_1}$ to the next layer ${\bf T_2}$. The 4 different configurations: in-phase and out-of-phase when ${\bf T_1}$ and ${\bf T_2}$ are parallel or anti-parallel and the two cases where the toroidal moments or anapoles are crisscrossed. In the last cases, the sum ${\bf T_1 + T_2}$ either points along a, or b. For each group, one example is represented on Fig.~4 of the manuscript where the pattern in the first plane, ${\bf T_1}$, is always the same. 
Another choice of ${\bf T_1}$ will give rise to another 4 patterns which will belong to each of the category discussed above. 
			
The magnetic intensity is proportional to the square of the magnetic structure factor $\rm |F_M|^2$ \cite{Review2011b}. For the LC model considered here, that corresponds to the out-of-plane magnetic intensity, $\rm I_{mag} \propto |F_M|^2 |M_c|^2$, measuring the orbital moment pointing along c*. 

We now estimate $\rm I_{mag}$ for the 4 configurations of Fig. 4. For the two first categories of Fig.~4, $\rm I_{mag}$ is proportional to $\rm |2 M_c sin (2 \pi x_0)|^2 |2 cos(\pi \frac{d}{c} L)|^2$ (for the in-phase or parallel moments configuration), or proportional to $\rm |2 M_c sin (2 \pi x_0)|^2 |2 sin(\pi \frac{d}{c} L)|^2$ (for the out-of-phase or antiparallel moments configuration). 

Finally, one can calculate $F_M$ for both crisscrossed configurations of Fig.~4. In the case of the sum ${\bf T_1 + T_2}$ of both anapoles is pointing along ${\bf a}$, we get:

\begin{widetext}
\begin{eqnarray}
F_M = 2 M_c [ 2 sin(\pi \frac{d}{c} L) sin(2 \pi x_0 H) cos(2 \pi x_0 K) + 2 cos(\pi \frac{d}{c} L) cos(2 \pi x_0 H) sin(2 \pi x_0 K) ] \\
I_{mag} [1,0,L] \propto |2 M_c sin (2 \pi x_0)|^2 |2 sin(\pi \frac{d}{c} L)|^2 \\
I_{mag} [0,1,L] \propto |2 M_c sin (2 \pi x_0)|^2 |2 cos(\pi \frac{d}{c} L)|^2
\end{eqnarray}
\end{widetext}

and in the case of the sum ${\bf T_1 + T_2}$ of both anapoles is pointing along ${\bf b}$:

\begin{widetext}
\begin{eqnarray}
F_M = 2 M_c [ 2 cos(\pi \frac{d}{c} L) sin(2 \pi x_0 H) cos(2 \pi x_0 K) + 2 sin(\pi \frac{d}{c} L) cos(2 \pi x_0 H) sin(2 \pi x_0 K) ] \\
\label{parab}
I_{mag} [1,0,L] \propto |2 M_c sin (2 \pi x_0)|^2 |2 cos(\pi \frac{d}{c} L)|^2 \\
\label{parab10L}
I_{mag} [0,1,L] \propto |2 M_c sin (2 \pi x_0)|^2 |2 sin(\pi \frac{d}{c} L)|^2
\label{parab01L}
\end{eqnarray}
\end{widetext}
				
These relations are given in table I of the manuscript for each direction. It is clear from these expressions that another choice of ${\bf T_1}$  in the lower layer will give rise to the same magnetic intensity for each of the category as it will only change the sign of $x_0$. 

One finally remarks that when the putative mirror symmetry is broken, it should be reflected in a change of the crystal structure with a distortion that should be measured either by neutron scattering or X-ray measurements. However, the distortion can be too weak to be observed. So far, no distortion of the orthorhombic crystal lattice of YBCO has been reported at T*. Interestingly, the broken mirror symmetry at T* has been also recently shown by the second harmonic optics results \cite{Hsieh2016b}.

\end{document}